\documentclass[12pt]{article}
\usepackage{epsfig}
\usepackage{cite}
\begin{document}

\begin{titlepage}

\hfill FTUV-11-0721

\hfill IFIC/11-37

\vspace{1.5cm}

\begin{center}
\ 
\\
{\bf\large  Heavy meson interquark potential }
\\
\date{ }
\vskip 0.70cm

P. Gonz\'alez, V. Mathieu
\\ and V. Vento

\vskip 0.30cm

{ \it Departamento de F\'{\i}sica Te\'orica -IFIC\\
Universidad de Valencia-CSIC \\
E-46100 Burjassot (Valencia), Spain.} \\ ({\small E-mail:
name.surname@uv.es}) 
\end{center}

\vskip 1cm \centerline{\bf Abstract}
The resolution of Dyson-Schwinger equations leads to the freezing of the QCD running coupling (effective charge) in the infrared, which is best understood as a dynamical generation of a gluon mass function, giving rise to a momentum dependence which is free from infrared divergences. We calculate the interquark static potential for heavy mesons by assuming that it is given by a massive One Gluon Exchange interaction and compare with phenomenologyical fits inspired by lattice QCD. We apply these potential forms to the description of quarkonia  and conclude that, even though some aspects of the confinement mechanism are absent in the Dyson-Schwinger formalism, the spectrum can be reasonably reproduced. We discuss possible explanations for this outcome.

 \vspace{1cm}

\noindent Pacs: 12.38.Gc, 12.38.Lg, 12.39.Pn, 14.40.Pq


\end{titlepage}

\section{Introduction}
The discovery of asymptotically free constituents of hadronic matter in deep
inelastic scattering experiments gave birth to Quantum Chromodynamics (QCD) 
\cite{Fritzsch:1973pi} as the accepted theory of strong
interactions. However, the most precise experimental data, the hadron
spectrum and hadron properties, have been obtained in the low-energy region. While
QCD explains asymptotic freedom \cite{Gross:1973id,Politzer:1973fx}, it has not been shown to
describe confinement of quarks and gluons, nor the realization of chiral symmetry. Therefore,
predicting low-energy properties of strongly interacting matter  represents a  theoretical challenge. 

The development of nonperturbative techniques are essential for undertsanding  QCD in this regime. Lattice gauge theory \cite{Wilson:1974sk,Creutz:1980zw} constitutes a nonperturbative regularisation scheme which allows numerical solutions of the theory describing properties of interacting  QCD matter   \cite{Creutz:2011hy}. The accuracy of lattice results has been tremendously improved  during the past decade with the availability of more powerful computers \cite{Sachrajda:2011tg,Hagler:2011zz}, and lattice results are considered in many instances the data  which other nonperturbative schemes should reproduce.

The approximate resolution of the Dyson-Schwinger (DS) equations  is another nonperturbative approach which has progressed considerably in the last ten years in great part due to the interplay between their findings and lattice results. It is a more analytical approach and  has led to a very appealing physical picture establishing that the QCD running coupling (effective charge) freezes in the deep infrared. This property can be best understood from the point of view of a dynamical gluon mass generation \cite{Cornwall:1982zr,Aguilar:2006gr}. 

The aim of this presentation is to investigate the consequences associated to the static interaction one can derive from this picture. For this purpose we calculate numerically the one gluon exchange (OGE) static potential deriving from the DS equations and we  compare it to phenomenological potentials whose shape has been inspired by lattice computations. The application of these potentials to the description of quarkonia is discussed.

The paper is written as follows, in the next section we describe some phenomenological potentials justified or motivated by lattice calculations. In Sec. 3 we  obtain a potential from the resolution of DS equations and  describe 
some of its characteristics. In Sec. 4 we make a comparative analysis of the different potentials in their application to quarkonia. Finally, in Sec. 5, we  summarize our main findings and discuss a possible interpretation of our results.

\section{Heavy quark dynamics from Lattice}

Soon after the discovery of the J/$\Psi$ mesons in $e^+e^-$ annihilation, the possibility of a nonrelativistic treatment of such states, in analogy to the positronium of electrodynamics, was suggested \cite{Appelquist:1975ya}. Quarkonia, i.e. mesonic states that contain two heavy constituent quarks, either charm or bottom, owe their name to this analogy. 

For sufficiently heavy quarks, it can be shown in lattice QCD, that the bound state problem becomes essentially nonrelativistic and the dynamics is controlled approximately by a Schr\"odinger equation with a static potential \cite{Bali:2000gf,Brambilla:2004jw}. Lattice calculations, without dynamical quarks (quenched), give rise to a  static potential of the form \cite{Bali:2000gf,Brambilla:2004jw} 
\begin{equation}
V(r)=-a/r+br.  
\label{Cornell}
\end{equation}%
containing the perturbative expected Coulomb term plus an additional linear term. 

The Cornell group, prior to the lattice QCD derivation, using $a$ and $b$ as parameters,  successfully applied this potential  to the phenomenological description of the low lying quarkonia states \cite{Eichten:1974af,Quigg:1979vr,Eichten:1979ms,Eichten:2007qx}
and therefore this potential function is known as the Cornell potential.

%%%%%%%%%%%%%%%%%%%%%%%% FIGURE 1 %%%%%%%%%%%%%%%%%%%%%%%%%%%%%%%%%%%%
%Fig 1. heavy meson potentials
%%%%%%%%%%%%%%%%%%%%%%%%%%%%%%%%%%%%%%%%%%%%%%%%%%%%%%%%%%%%%%%%%%%%%
\begin{figure}[t]
\begin{center}
\includegraphics[scale=1]{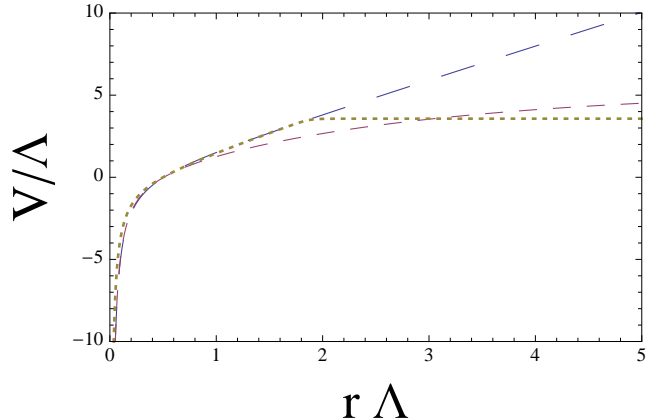}
\caption{We show the three potentials described in the text : Cornell (solid), Aachen (dashed) 
and SESAM (dotted). For the Cornell   and Aachen potentials the parameters have been chosen phenomenologically to provide a reasonable 
description of the spectrum and their values are given in Sec. 4. The parameters of the SESAM potential are chosen to reproduce the shape of the Cornell potential at short distances.}
\label{latticepot}
\end{center}
\end{figure}
%%%%%%%%%%%%%%%%%%%%%%%%%%%%%%%%%%%%%%%%%%%%%%%%%%%%%%%%

Another static interquark potential shape has been guessed by performing unquenched lattice QCD calculations.  
Using Kogut-Susskind fermions and adopting the lattice spacing from the $\rho$
 mass measurements a potential with the following structure, 
\begin{equation}
V(r) = (-\alpha/r + \beta r)\, \left(\frac{1-e^{-\gamma r}}{\gamma r}\right),
\label{latticeB}
\end{equation}
has been shown to describe these lattice results\cite{Born:1989iv}.
This parameterization shows screening, associated to quarks loops, which has been cast in the form of the additional 
factor multiplying the Cornell potential, leading at large values of $r$ to a constant $\beta/\gamma$.  
Henceforth we shall call this shape the Aachen potential and we shall choose its parameters to provide a reasonable fit
to the spectrum.

More recent unquenched lattice calculations have been performed by the SESAM
collaboration \cite{Bali:2005fu}. The mixing and the transition (string breaking)
between $q\overline{q}$ and $M\overline{M}$ ($M$ representing a meson) components have
been studied. A correlation matrix for this two-level problem has been
considered, the matrix elements incorporating light quark propagators into
the standard Wilson loop. Compelling evidence, both for explicit mixing and
string breaking, has been demonstrated so that for $r\geq r_{c}>1$ fm ($r_{c}$
representing the string breaking distance) the $M\overline{M}$ configuration becomes
energetically favorable. The static $q\overline{q}$ potential exhibits
screening and saturates at twice the mass of the $M$ meson. An approximate
parametrization of it for $r\leq r_{c}$ has been proposed,

\begin{equation}
V(r)=m_{M}+m_{\overline{M}}+g(r)G(r)+C
\end{equation}
where
\begin{eqnarray}
g(r) & =&\frac{1}{2} -\frac{1}{\pi } \arctan \left[ d(r-r_{c})\right]  \nonumber\\
G(r) & = &-e\,\left(\frac{1}{r}-\frac{1}{r_{c}}\right)+\sigma (r-r_{c})  \nonumber
\end{eqnarray}
with $C,$ $r_{c},$ $d,$ $e$ and $\sigma $ obtained from lattice results.
This parameterization has not the correct long distance limit 
($m_{M}+m_{\overline{M}})$, and differs little from the Cornell potential
for $r<0.75$ fm.

We show the three potentials  in Fig. \ref{latticepot} for comparison\footnote{The spin-dependent corrections to the potentials
above can also be derived from lattice QCD  \cite{Koma:2006fw}. We do not consider them here.}.  We use in what follows the first two, which contain the two aspects we want to emphasize in here, linear confinement and screening.

\section{Heavy quark dynamics from Dyson-Schwinger Equations}

Infrared finite solutions for the gluon propagator of quenched QCD are obtained from the
gauge-invariant nonlinear Dyson-Schwinger equations formulated in the Feynman gauge of
the background field method. These solutions may be fitted using a massive propagator
\cite{Cornwall:1982zr,Aguilar:2006gr}. 
 Even though  the  gluon is massless  at the  level  of the  fundamental  QCD Lagrangian, and  remains massless to all order in perturbation theory, the nonperturbative QCD
dynamics  generates  an  effective,  momentum-dependent  mass,  without affecting    the   local    $SU(3)_c$   invariance,    which   remains
intact. 

The gluon mass generation is a purely nonperturbative effect associated with 
the existence of 
infrared finite solutions for the gluon propagator, $\Delta (q^2)$,
i.e. solutions with  $\Delta^{-1}(0) > 0$.
Such solutions may  
be  fitted  by   a  ``massive''  euclidean propagator  of   the form \cite{Cornwall:1982zr,Aguilar:2008xm} 

\begin{equation}
 \Delta(q^2) = \frac{1}{q^2  +  m^2(q^2)},
 \label{propagator}
 \end{equation}
where $m^2(q^2)$  depends nontrivially  on the momentum  transfer $q^2$.
One physically motivated possibility, which we shall use in here, is  the so called logarithmic mass running, which is defined by

\begin{equation}
m^2 (q^2)= m^2_0\left[\ln\left(\frac{q^2 + \rho m_0^2}{\Lambda^2}\right)
\bigg/\ln\left(\frac{\rho m_0^2}{\Lambda^2}\right)\right]^{-1 -\delta},
\label{rmass}
\end{equation}
where $m_0, \rho$ and  $\delta$ are parameters whose values are chosen to fit the lattice propagator and $\Lambda$ is the $QCD$ scale.
Note that in the limit $q^2\to 0$ one obtains $m^2(0)=m^2_0$, giving meaning to $m_0$.

Because of the presence of this dynamical gluon mass the strong effective
charge extracted from these solutions freezes at a finite value, giving rise to an infrared fixed
point for QCD \cite{Cornwall:1982zr,Aguilar:2006gr}. This  nonperturbative  generalization  of $\alpha(q^2)$,
the  QCD  running  coupling, comes in the form
\begin{equation}
a(q^2) = \left[\beta_0 \ln \left(\frac{q^2 +\rho \, m^2(q^2)}{\Lambda^2}\right)\right]^{-1} ,
\label{alphalog}
\end{equation}
where  $a =\frac{\alpha}{4 \pi}$ and we take $\beta_0 = 11 - 2 n_f/3$ where $n_f$ is the number of flavors. Note that its zero gluon mass limit leads to the LO perturbative 
coupling constant momentum dependence.
The $m(q^2)$ in the argument of the logarithm 
tames  the   Landau pole, and $a(q^2)$ freezes 
at a  finite value in the IR, namely  
\mbox{$a^{-1}(0)= \beta_0 \ln (\rho \, m^2(0)/\Lambda^2)$} .

\begin{figure}[htb]
\begin{center}
\includegraphics[scale=0.4]{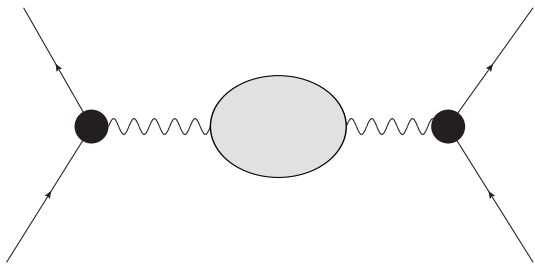}
\caption{One gluon exchange contribution to the potential.}
\label{oge}
\end{center}
\end{figure}

Let us construct a simple potential model where the main source of dynamics is the One Gluon Exchange potential (see Fig. \ref{oge}) with the propagator and coupling defined by Eqs. \ref{propagator}, \ref{rmass} and \ref{alphalog}.

%%%%%%%%%%%%%%%%%%%%%%%% FIGURE 2 %%%%%%%%%%%%%%%%%%%%%%%%%%%%%%%%%%%%
%      Fig 2. One-gluon exchange
%%%%%%%%%%%%%%%%%%%%%%%%%%%%%%%%%%%%%%%%%%%%%%%%%%%%%%%%%%%%%%%%%%%%%
%%%%%%%%%%%%%%%%%%%%%%%%%%%%%%%%%%%%%%%%%%%%%%%%%%%%%%%%

 The potential between static charges is related to the Fourier transform of the time-time component of the  full gluon propagator as,

\begin{equation}
V({\mathbf r}) = - \frac{2 C_F } {\pi} \int d^3 {\mathbf q }\,a(\mathbf q^2)\,\Delta_{00}({\mathbf q})e^{i{\mathbf {q \cdot r}}},	
\label{pot}
\end{equation}
where $C_F$ is the Casimir eigenvalue of the fundamental representation of SU(3) [$C_F=4/3$],
the bold terms, ${\mathbf q}$ and ${\mathbf r}$, are 3-vectors and $\Delta_{00}({\mathbf q})$ is the zero-zero component of
the gluon propagator in the momentum configuration. 

\begin{figure}[htb]
\begin{center}
\includegraphics[scale=1.0]{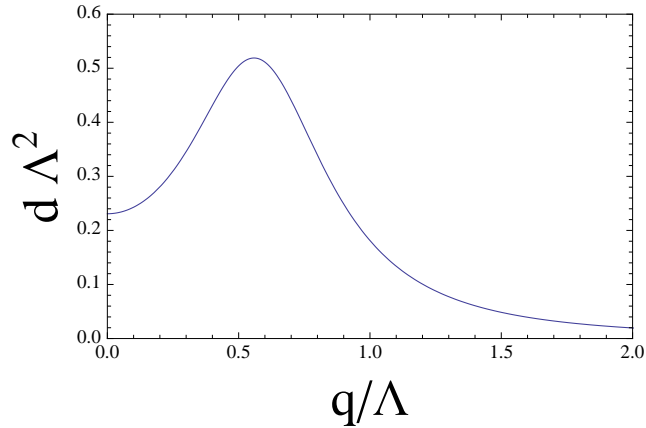}
\caption{Numerical solution for the RG invariant product $d$ for $m_0 = 360$ MeV, $\rho =1$, $\Lambda = 300$ MeV  and $n_f = 4$. We plot the adimensional function $d \Lambda^2$ as a function of $\mathbf q/\Lambda$.}
\label{d}
\end{center}
\end{figure}

Using the above equations the expression $a(\mathbf q^2)\,\Delta_{00}({\mathbf q})$ in the integrand of Eq.\ref{pot} becomes
\begin{equation}
d(\mathbf q^2) = \,\frac{a (\mathbf q^2)}{\mathbf q^2 +m^2(\mathbf q^2)} 
\label{prop}
\end{equation}
with the dynamical mass $m^2(\mathbf q^2)$  and the nonperturbative coupling constant $a (\mathbf q^2)$ defined by Eqs. \ref{rmass} and \ref{alphalog} . This particular combination of propagator and coupling constant is renormalization group invariant. The structure of $d (\mathbf q^2)$ is shown in Fig. \ref{d}.

%%%%%%%%%%%%%%%%%%%%%%%% FIGURE 3 %%%%%%%%%%%%%%%%%%%%%%%%%%%%%%%%%%%% 
%     The RG momentum potential
%%%%%%%%%%%%%%%%%%%%%%%%%%%%%%%%%%%%%%%%%%%%%%%%%%%%%%%%%%%%%%%%%%%%%
%%%%%%%%%%%%%%%%%%%%%%%%%%%%%%%%%%%%%%%%%%%%%%%%%%%%%%%%

In spherical coordinates Eq.\ref{pot} becomes,

\begin{equation}
V({\mathbf r})= -\frac{2 \,C_F }{\pi}\int_{0}^{2\pi}\!\!\!d\phi \int_{0}^{\infty}\!\!\!d{\mathbf{|q|\,q^2}}\,a(\mathbf q^2)\,\Delta({\mathbf q ^2}) \int_{-1}^{1}\!\!\!d (\cos \theta) \,e^{i{\mathbf {|q||r|\cos \theta}}}.	
\end{equation}
By performing the angular integral we obtain,
\begin{equation}
V({\mathbf r})=	- \frac{8\, C_{F}}{|\mathbf r|}\int_{0}^{\infty}\!\!\! d {|\mathbf q}|\;{|\mathbf q}|\,a( \mathbf q^2)\,\Delta({\mathbf q^2})
\sin({|{\mathbf q}||{\mathbf r}|}).
\label{fourier}
\end{equation}
%

%%%%%%%%%%%%%%%%%%%%%%%% FIGURE 4 %%%%%%%%%%%%%%%%%%%%%%%%%%%%%%%%%%%%
%     The static potential without Sommer subtraction
%%%%%%%%%%%%%%%%%%%%%%%%%%%%%%%%%%%%%%%%%%%%%%%%%%%%%%%%%%%%%%%%%%%%%
\begin{figure}[h]
\begin{center}
\includegraphics[scale=0.85]{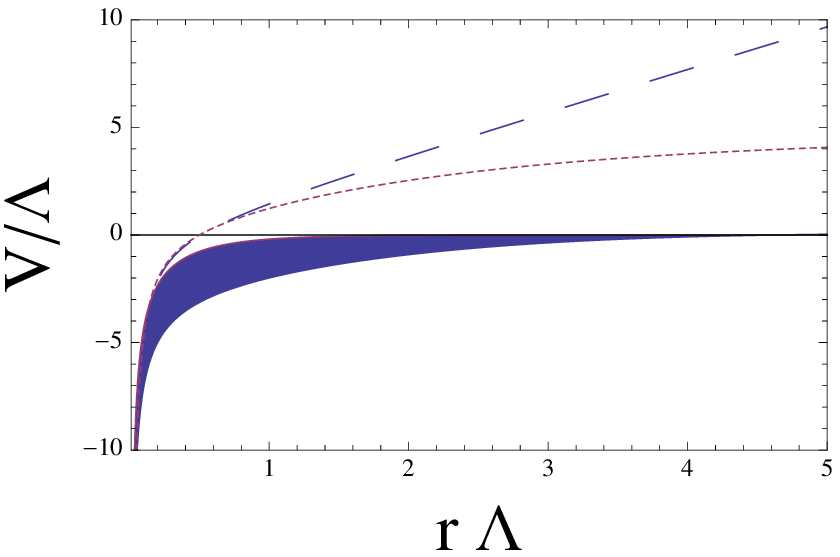}
\hspace{0.75cm}
\includegraphics[scale=0.85]{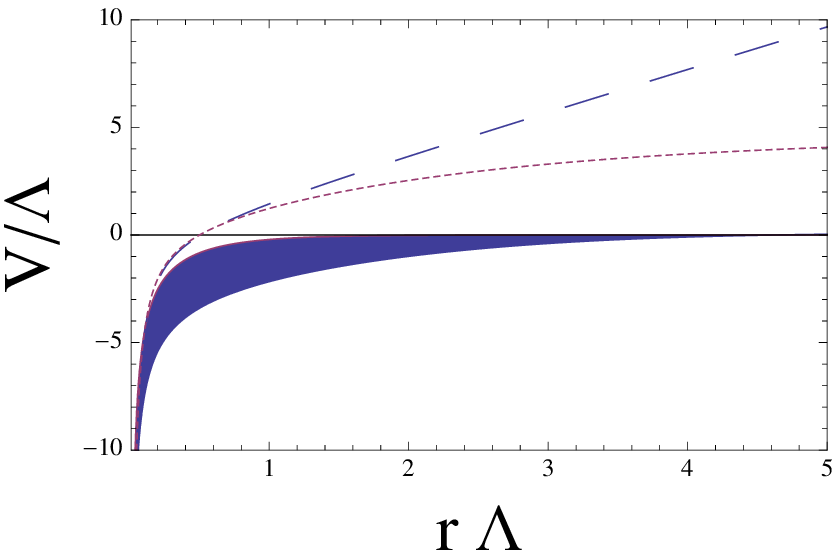}
\caption{We show the  potential  obtained from the DS equations. The shaded range corresponds  to $m_0 \sim 360-480 $MeV and $\rho \sim 1-4 $. The figure on the left corresponds to 4 flavors( $\beta_0 =25/3$). The figure on the right corresponds to 5 flavors ($\beta_0 = 23/3$). For comparison we plot  the Cornell (dashed) and Aachen (dotted) potentials.}
\label{DSEpot}
\end{center}
\end{figure}
%%%%%%%%%%%%%%%%%%%%%%%%%%%%%%%%%%%%%%%%%%%%%%%%%%%%%%%%

The complete QCD potential involves multigluon exchanges, but we assume that the leading nonperturbative term will be sufficient for the description of the spectrum.

%%%%%%%%%%%%%%%%%%%%%%%% FIGURE 5 %%%%%%%%%%%%%%%%%%%%%%%%%%%%%%%%%%%%
%     The static potential with Sommer subtraction
%%%%%%%%%%%%%%%%%%%%%%%%%%%%%%%%%%%%%%%%%%%%%%%%%%%%%%%%%%%%%%%%%%%%%
\begin{figure}[htb]
\begin{center}
\includegraphics[scale=1.0]{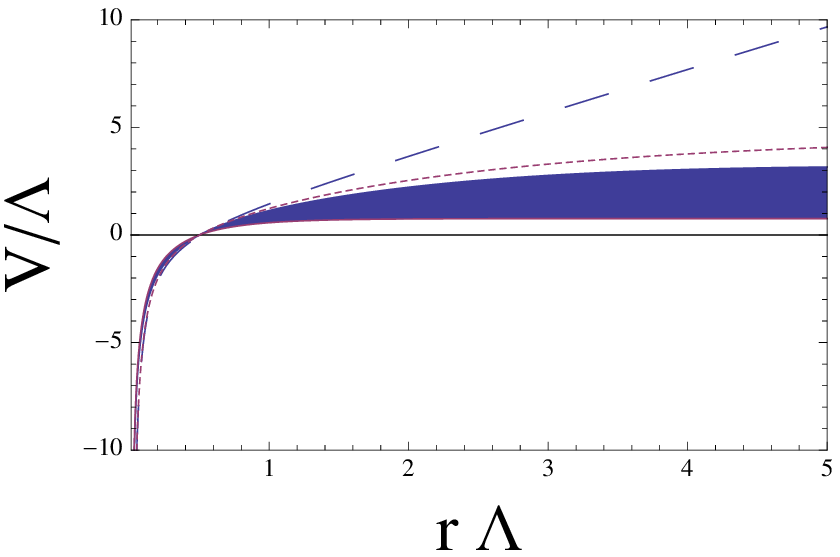}
\caption{We show the DS potential after the Sommer subtraction, for $n_f=4$ and the same range of parameters as in Fig. \ref{DSEpot}. For comparison we plot  the Cornell (dashed) and Aachen (dotted) potentials.}
\label{DSEpotsom}
\end{center}
\end{figure}
%%%%%%%%%%%%%%%%%%%%%%%%%%%%%%%%%%%%%%%%%%%%%%%%%%%%%%%%

In Fig. \ref{DSEpot} we show  the potential (up to a constant) derived from  Eq. \ref{fourier}, i.e. from the DS equations with the definitions in Eqs.  \ref{rmass} and \ref{alphalog}.  We have chosen the following range of parameters: $m_0 \sim 360-480$ MeV, $\rho=1 - 4$, $\delta= 1/11$ \cite{Cornwall:1982zr,Aguilar:2007ie,Aguilar:2009nf}. The value of $\Lambda$ has been taken to be $300$ MeV. In order to adjust the behavior at the origin to the data we have  used $\beta_0$ corresponding to $4$ and $5$ flavors. To do this appropriately one should introduce the running of the quark masses, which are at present not well known. However, since asymptotically  the masses run to zero, our way of proceeding achieves the correct strength of the potential at low $r$. The potential describes  well  the low radial behavior, by construction, and flattens at large $r$ going asymptotically to zero.  

We should remove from the above calculation the additive infinite self-energy contribution associated  with the static sources \cite{Bali:2000gf}. In lattice QCD this is done normalizing the potential such that $V(r_0)=0$ where $r_0$ is the Sommer scale \cite{Sommer:1993ce} . We proceed in the same way but take the subtraction point at the zero  of the phenomenological potentials, which happens to  be at $r_0 \sim 0.35 fm$ (see next section). The result of this procedure is shown in Fig. \ref{DSEpotsom}. In this way we increase the value of the potential without changing its shape. We call the result of this construction the Dyson-Schwinger (DS) potential . We obtain in this way a potential which resembles the Aachen and not the Cornell potential.
 
It should be noted that the shape of the DS potential does not vary considerably when we change the parameters within the expected theoretical range. There is no way to reproduce the large $r$ behavior of the Cornell potential, instead the DS potential flattens and becomes asymptotically constant, similarly to the screened potentials. However, it should not! The approximations used to find the solution to the DS equations do not contain quark loops and therefore they incorporate  no mechanism for screening, i.e. a mechanism  derivable from the breaking of the string \cite{Born:1989iv,Kratochvila:2003zj}. The truncation in the DS set of equations and the absence of multigluon exchanges might be responsible for the missing linear rise at large $r$.

\section{Quarkonia description by static potentials}

The Cornell potential was introduced for phenomenological reasons predating the  complete
lattice derivation. From the spin-averaged quarkonia spectra it was evident  that the 
underlying potential could not be purely Coulomb type \cite{Eichten:1974af} . 
Therefore, the potential  was implemented by a sum of the perturbative expectation 
plus an additional linear term, recall Eq. \ref{Cornell}.  Aiming at a universal treatment for 
charmonium ($c\overline{c}$) and bottomonium ($b\overline{b}$) the Cornell 
potential has been used with the same values of the parameters, $a$ and $b$, in both cases, for 
the quark masses, $m_{c}$ and $m_{b}$, respectively. A typical range of values providing a 
reasonable fit to the masses of the low lying states (up to 1.0 GeV
excitation energy) is $a\sim 0.51-0.52$ and $\sqrt{b}\sim 412-427$ MeV  
\cite{Quigg:1979vr,Eichten:1979ms, Eichten:2007qx}.
An example of such spectral fit for charmonium is provided in 
Table 1 where we have chosen $\sqrt{b}=427$ MeV and $a=0.52$ with $%
m_{c}=1350 $ MeV and compared the results with masses of experimental
resonances having well established $J^{PC}$ quantum numbers, most of them
with $J^{PC}=1^{--}$ coming from ISR (Initial State Radiation) processes.
The Cornell model provides a good fit to the lower states
(at most 30 MeV difference with data) but it cannot accommodate all the
known higher energy resonances but only some of them (we use charmonium
instead of bottomonium to clarify the effect). For instance $\psi
(4040),$ $\psi (4160)$ and $\psi (4415)$ maybe assigned to the $3s,$ $2d$
and $4s$ states respectively. Then other two resonances, cataloged in the
Particle Data Group Review \cite{Nakamura:2010zzi} as $X(4260)$ and $X(4360),$ cannot
be fitted. 

The screened potentials introduced initially to fit lattice data have been used
phenomenologically \cite{Ding:1995he,Gonzalez:2003gx}.
However, the applicability of screened potentials to the spectral description
has been a matter of debate \cite{Swanson:2005rc} since their use is not justified
above the meson-meson string breaking threshold. We  assume that in the
case of quarkonia there is an effective string breaking threshold sufficiently high in
energy,  to allow for a description of 
the known spectrum in terms of screened potentials. For the sake of comparison 
the results from the Aachen potential for $m_{c}=1400$ MeV and the same values 
for the corresponding Cornell parameters $\alpha=a,$ $\beta =b$ are presented in Table 1. 
The value of $\gamma =0.38$ fm$^{-1}$ is fixed to get a reasonable spectral fit.
It is important to emphasize that the values of the parameters of the Aachen 
potential extracted from lattice data \cite{Born:1989iv}  can only  give a reasonable 
description of the masses of the low lying quarkonia.

\begin{table}[tbp]
\begin{center}
\begin{tabular}{|cccc|c|}
\hline
$n_{r}L$ & $M_{Cornell}$ & $M_{Aachen}$ & $M_{PDG}$ & $M_{DS}$ \\ \hline
& MeV & MeV & MeV & MeV \\ \hline
$1s$ & 3069 & 3143 & $3096.916\pm 0.011$ & 3151 \\ 
$2s$ & 3688 & 3665 & $3686.09\pm 0.04$ & 3660 \\ 
$1d$ & 3806 & 3775 & $3772.92\pm 0.35$ & 3761 \\ 
$3s$ & 4147 & 4001 & $4039\pm 1$ & 4004 \\ 
$2d$ & 4228 & 4072 & $4153\pm 3$ & 4070 \\ 
$4s$ & 4539 & 4255 & $4263_{-9}^{+8}$ & 4273 \\ 
$3d$ & 4601 & 4306 & $4361\pm 9\pm 9$ & 4321 \\ 
$5s$ & 4829 & 4564 & $4421\pm 4$ & 4487 \\ 
$4d$ & 4879 & 4609 &  & 4526 \\ 
$6s$ & 5218 & 4629 & $4664\pm 11\pm 5$ & 4651 \\ \hline\hline
$5d$ & 5260 & 4663 &  & 4718 \\ \hline\hline
$1p$ & 3502 & 3527 & $3525.3\pm 0.2$ & 3515 \\ 
$2p$ & 3983 & 3894 &  & 3886 \\ \hline
\end{tabular}%
\caption{Calculated charmonium masses, $M_{Cornell}$, $M_{Aachen}$ and $%
M_{DS}$ from the Cornell, Aachen and DS potentials. For Cornell
and Aachen $a=\alpha = 0.52$ and $\sqrt{b}=\sqrt{\beta }=427$ MeV. The remaining parameter in the Aachen potential
has been chosen to be $\gamma =0.38$ fm$^{-1}$. For DS $\rho =1$ and $m_{0}=345.7$ MeV. 
The charm masses are $m_{c}=1350$
MeV for Cornell and $m_{c}=1400$ MeV for Aachen and DS. Masses for
experimental candidates, $M_{PDG},$ have been taken from \protect\cite{Nakamura:2010zzi}%
. For $p$ waves we quote the centroid of the $np_{0}$, $np_{1}$ and $np_{2}$
states. }
\label{Table}
\end{center}
\end{table}

One should keep in mind that these potentials do not contain
spin-dependent terms which makes them reliable only when these terms do not
play a major role. We use here the conventional approximations, e.g. we consider
that these potentials should fit better the spin triplet states and take the centroids of $p$
states as data for comparison with our results.

The main difference between the Cornell and Aachen potentials refers to the
description of the higher excited states. The Aachen potential may allow a one 
to one assignment of the calculated states to the data.

The similarity of the DS potential (for certain parameter sets) to the Aachen potential  as shown in Fig. \ref{DSEpotsom} motivates the exercise of fitting the spectrum with the DS potential. As can bee seen in Table 1 this can be achieved for a set of parameters which is close to the expected theoretical  range. A low value of the mass, $m_0= 345.7$ MeV, is necessary since only with a value close to $\Lambda$ one gets sufficient strength to achieve, after Sommer subtraction,  an asymptotic behavior close to the Aachen potential used in our spectral fit.

Let us discuss the limitations of the parameters used in the fit. The $\beta_0$ expression is fixed by QCD. $\beta_0$  is the  leading order coefficient of the beta function of the theory, which is scheme independent; the small difference in value implied by the choice of $n_f$ has almost no effect on the results.  $\Lambda$ is a scale fixed by dimensional transmutation. We take $300$ MeV and keep it fixed. The values obtained in experimental  fits range from $250-300$ MeV but we have not even used that liberty.  The $\delta$ value and the variations in the values of $\rho$ and $m_0$ are within the limits provided  by the DS calculations, in particular $\Lambda < m_0 < 2 \Lambda$ [10,21].

The fit of the spectrum is strongly dependent on the detailed shape of the potential as determined by the mass function and coupling constant, Eqs. \ref{rmass} and \ref{alphalog} respectively.
Recall that the mass function is a fitting function to the lattice results for the propagator \cite{Aguilar:2008xm}. The quality of the spectral fit seems to point out, that the DS potential may qualitatively be a good description of the dynamics.

\section{Discussion and Conclusions}

\begin{figure}[htb]
\begin{center}
\includegraphics[scale=0.5]{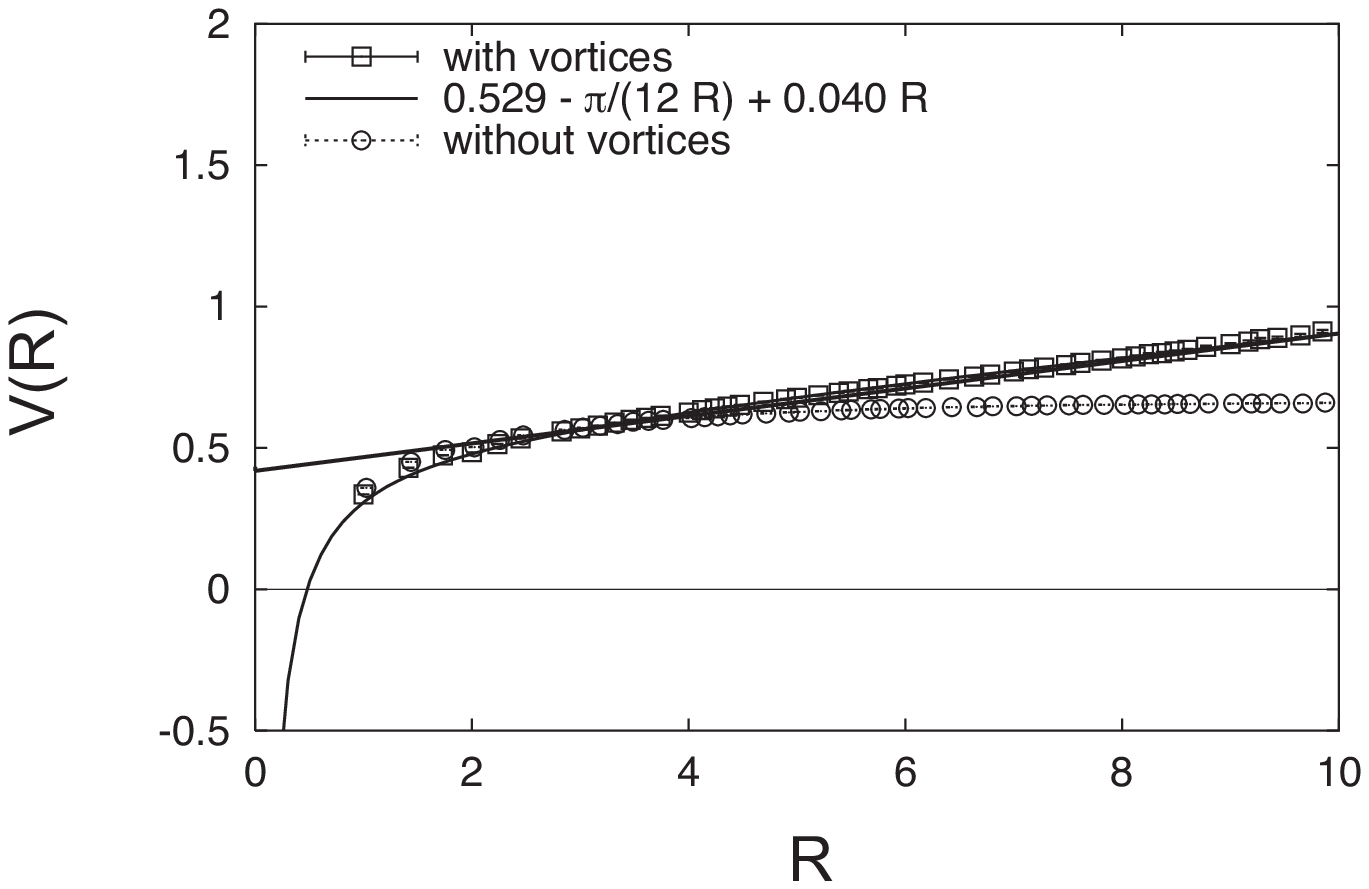}
\vspace{0.1cm}\hspace{0.5cm}
\includegraphics[scale=0.82]{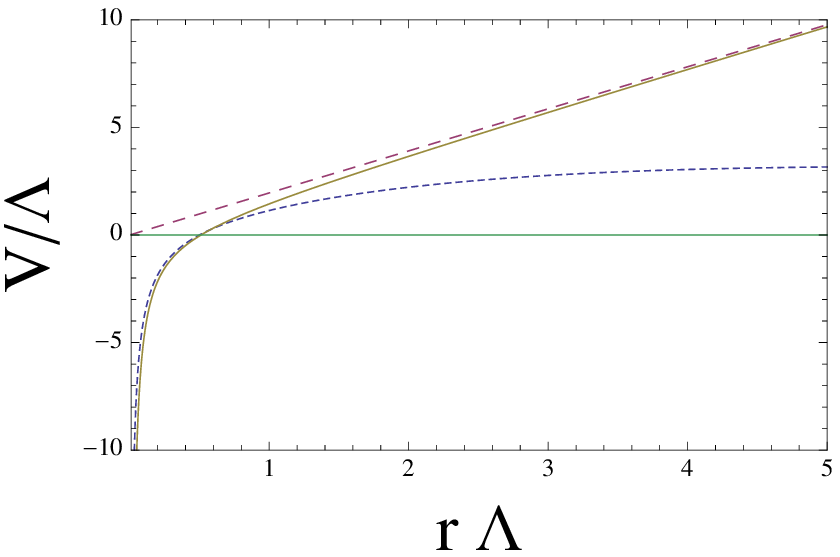}
\caption{ The  figure on the left corresponds to the calculation of ref.\cite{Greensite:2003xf}. In the figure on the right the dotted line corresponds to the DS potential  for  $n_f=4$ and $m_0=360$ MeV and $\rho=1$. The dashed line corresponds to the linear piece of the Cornell potential. The full Cornell potential is drawn (solid line) for comparison.}
\label{Coulomb}
\end{center}
\end{figure}

In order to interpret the above results we have to resort again to lattice calculations. In ref. \cite{Greensite:2003xf} the authors investigated the origin of the long range linearly rising potential. Using a mechanism described in ref. \cite{deForcrand:1999ms} they were able to subtract the contribution of the center vortices leading to a flat potential as shown in Fig. \ref{Coulomb} on the left. On the right of Fig. \ref{Coulomb} we show the DS potential after Sommer subtraction and the linear rise of the Cornell potential. It is quite apparent that the figure resembles the one on the left. Therefore we may tentatively conclude that the DS potential contains the physics associated with the approximate gluon interaction but does not contain the physics of the confinement mechanism. 
We do  not advocate  any mechanism for confinement in QCD, we only adhere to the fact that the confinement mechanism, whatever it be, is the mechanism behind the rising potential in the quenched approximation.

%%%%%%%%%%%%%%%%%%%%%%%% FIGURE 7 %%%%%%%%%%%%%%%%%%%%%%%%%%%%%%%%%%%%
%     Comparison between DS and Cornell through lattice
%%%%%%%%%%%%%%%%%%%%%%%%%%%%%%%%%%%%%%%%%%%%%%%%%%%%%%%%%%%%%%%%%%%%%
%%%%%%%%%%%%%%%%%%%%%%%%%%%%%%%%%%%%%%%%%%%%%%%%%%%%%%%%

A  surprising result of our calculation is the actual similarity of the DS potential to the Aachen potential as shown in Fig. \ref{DSEpotsom}. The latter arises due to the breaking of the string and is represented by a screened potential \cite{Born:1989iv}.  The similarity is astonishing more so since we have used conventional values for $\rho = 1, \delta= 1/11, \Lambda = 300 $ MeV and only varied  $m_0$, which becomes for the best fit $m_0= 345.7$  MeV within the expected range $ \Lambda < m_0 < 2 \Lambda$ \cite{Aguilar:2006gr,Aguilar:2008xm}.

Not aiming at such precision  if we fit the upper extreme of the
shaded region, where the DS potential is defined by $m_0=360$ MeV and $\rho =1$,  to the Aachen potential shape function we get  $\alpha=0.40$, $ \sqrt{\beta} = 412$ MeV
and $\gamma =0.53$ fm$^{-1}$. The other extreme set of parameters for
the DS potential, the lower edge of the shaded region, $m_0=480$ MeV and 
$\rho =4$, leads to $\alpha =0.37,\sqrt{\beta} =361$ MeV
and $\gamma =1.81$ fm$^{-1}$. 
 It is clear then, that  the main difference between the  parametrizations arises due to the value of the different screening ranges, which in the Aachen  potential  is controlled by  $\gamma$.
 
Is this similarity accidental or does string breaking imply a dilution of the confinement mechanism associated with multigluon exchanges and/or higher order truncation schemes ?
Does the Sommer subtraction introduce the energy scale of string breaking ?  More research
needs to be done in order to understand the confinement and string breaking mechanisms though
our investigation hints a possible scenario.

In conclusion, we have calculated the OGE potential associated to the approximate resolution of the Dyson-Schwinger equations for the gluon propagator. The low $r$ behavior is determined perturbatively. The large $r$ behavior is certainly nonperturbative. The Sommer procedure, to avoid self-energy effects of the static charges,  leads to a potential which is not negative at large $r$. The approximate resolution of the DS equations  is not able to reproduce, at the level of the OGE, the Cornell potential. The DS equations  together with the Sommer subtraction is close to the Aachen potential that  contains a string breaking mechanism. The fact that we have to push the parameters to the limit of the allowed region to reproduce the spectrum maybe due to the fact that we have not treated the quark mass terms appropriately. 

The potential derived from the DS equations  might contain most of the dynamics associated with the interquark interaction apart from a non trivial constant which might be related, in the unquenched calculation, to the confinement mechanism.

\section*{Acknowledgement}
We would like to thank very instructive conversations with Arlene Aguilar, Daniele Binosi and Joannis Papavassiliou. VV would like to acknowledge  clarifying conversations with Stefan Olejnik regarding ref. \cite{Greensite:2003xf}, with Nora Brambilla  regarding pNRQCD and the static lattice potential and discussions with Simonetta Liuti about the consequences of our results.
We thank the authors of JaxoDraw  for making drawing diagrams an
easy task \cite{Binosi:2003yf}.   This work  has been supported  by HadronPhysics2,  by  MICINN (Spain) grant  FPA2010-21750-C02-01,   AIC10-D-000598 and by GVPrometeo2009/129. VM has been supported by a post-doctoral grant from CPAN and by a contract from HadronPhysics2.

\end{document}